\documentclass[
 reprint,
 amsmath,amssymb,
 aps,
]{revtex4-2}
\usepackage{graphicx}
\usepackage{dcolumn}
\usepackage{bm}
\usepackage{braket}
\usepackage{here}
\usepackage{color}
\newcommand{\qq}{\Braket{\bar{q}q}}
\newcommand{\etap}{\eta'}

\newcommand{\C}[1]{{}^{#1}{\rm C}}

\begin{document}
\title{\boldmath{Excitation Spectra of the $\C{12}(p,d)$ Reaction near the $\eta'$-Meson Emission Threshold
  Measured in Coincidence with High-Momentum Protons}}
\author{R.~Sekiya$^{1,2,3}$}
\author{K.~Itahashi$^{2,3,4}$}
\email{Email: itahashi@a.riken.jp}
\author{Y.~K.~Tanaka$^{5,6}$}
\email{Email: yoshiki.tanaka@a.riken.jp}
\author{S.~Hirenzaki$^{7}$}
\author{N.~Ikeno$^{8}$}
\author{V.~Metag$^{9}$}
\author{M.~Nanova$^{9}$}
\author{J.~Yamagata-Sekihara$^{10}$}
\author{V.~Drozd$^{6,11}$}
\author{H.~Ekawa$^{5}$}
\author{H.~Geissel$^{6,9}$}
\email{Deceased}
\author{E.~Haettner$^{6}$}
\author{A.~Kasagi$^{5,12}$}
\author{E.~Liu$^{5,13}$}
\author{M.~Nakagawa$^{5}$}
\author{S.~Purushothaman$^{6}$}
\author{C.~Rappold$^{14}$}
\author{T. R.~Saito$^{5,6,15}$}
\author{H.~Alibrahim Alfaki$^{6}$}
\author{F.~Amjad$^{6}$}
\author{M.~Armstrong$^{6}$}
\author{K.-H.~Behr$^{6}$}
\author{J.~Benlliure$^{16}$}
\author{Z.~Brencic$^{17,18}$}
\author{T.~Dickel$^{6,9}$}
\author{S.~Dubey$^{6}$}
\author{S.~Escrig$^{14}$}
\author{M.~Feijoo-Font\'an$^{16}$}
\author{H.~Fujioka$^{19}$}
\author{Y.~Gao$^{5,13}$}
\author{F.~Goldenbaum$^{20}$}
\author{A.~Gra\~{n}a Gonz\'{a}lez$^{16}$}
\author{M. N.~Harakeh$^{11}$}
\author{Y.~He$^{5,15}$}
\author{H.~Heggen$^{6}$}
\author{C.~Hornung$^{6}$}
\author{N.~Hubbard$^{6,21}$}
\author{M.~Iwasaki$^{2,3}$}
\author{N.~Kalantar-Nayestanaki$^{11}$}
\author{M.~Kavatsyuk$^{11}$}
\author{E.~Kazantseva$^{6}$}
\author{A.~Khreptak$^{22,23}$}
\author{B.~Kindler$^{6}$}
\author{H.~Kollmus$^{6}$}
\author{D.~Kostyleva$^{6}$}
\author{S.~Kraft-Bermuth$^{24}$}
\author{N.~Kurz$^{6}$}
\author{B.~Lommel$^{6}$}
\author{S.~Minami$^{6}$}
\author{D. J.~Morrissey$^{25}$}
\author{P.~Moskal$^{22}$}
\author{I.~Mukha$^{6}$}
\author{C.~Nociforo$^{6}$}
\author{H.~J.~Ong$^{13}$}
\author{S.~Pietri$^{6}$}
\author{E.~Rocco$^{6}$}
\author{J.~L.~Rodr\'{i}guez-S\'{a}nchez$^{16,26}$}
\author{P.~Roy$^{6,27}$}
\author{R.~Ruber$^{28}$}
\author{S.~Schadmand$^{6,20}$}
\author{C.~Scheidenberger$^{6,9,29}$}
\author{P.~Schwarz$^{6}$}
\author{V.~Serdyuk$^{20}$}
\author{M.~Skurzok$^{22}$}
\author{B.~Streicher$^{6}$}
\author{K.~Suzuki$^{6,20}$}
\author{B.~Szczepanczyk$^{6}$}
\author{X.~Tang$^{13}$}
\author{N.~Tortorelli$^{6}$}
\author{M.~Vencelj$^{17,18}$}
\author{T.~Weber$^{6}$}
\author{H.~Weick$^{6}$}
\author{M.~Will$^{6}$}
\author{K.~Wimmer$^{6}$}
\author{A.~Yamamoto$^{30}$}
\author{A.~Yanai$^{5,31}$}
\author{J.~Zhao$^{6,32}$}
 
\affiliation{\it {}$^{1}$
 Division of Physics and Astronomy, Kyoto University, 606--8502 Kyoto, Japan}

\affiliation{\it {}$^{2}$
 Meson Science Laboratory, RIKEN, Wako, 351--0198 Saitama, Japan}

\affiliation{\it {}$^{3}$
 Nishina Center for Accelerator--Based Science, RIKEN, Wako, 351--0198 Saitama, Japan}

\affiliation{\it {}$^{4}$
 Department of Physics, The University of Osaka, Toyonaka, 560--0043 Osaka, Japan}

\affiliation{\it {}$^{5}$
 High Energy Nuclear Physics Laboratory, RIKEN, Wako, 351--0198 Saitama, Japan}

\affiliation{\it {}$^{6}$
 GSI Helmholtzzentrum f\"{u}r Schwerionenforschung GmbH, 64291 Darmstadt, Germany}

\affiliation{\it {}$^{7}$
 Department of Physics, Nara Women's University, 630--8506 Nara, Japan}

\affiliation{\it {}$^{8}$
 Graduate School of Maritime Sciences, Kobe University, 658--0022 Hyogo, Japan}

\affiliation{\it {}$^{9}$
 II. Physikalisches Institut, Universit\"{a}t Gie\ss en, 35392 Gie\ss en, Germany}

\affiliation{\it {}$^{10}$
 Department of Physics, Kyoto Sangyo University, 603--8555 Kyoto, Japan}

\affiliation{\it {}$^{11}$
 ESRIG, University of Groningen, 9747 AA Groningen, The Netherlands}

\affiliation{\it {}$^{12}$
 Department of Engineering, Gifu University, 501--1193 Gifu, Japan}

\affiliation{\it {}$^{13}$
 Institute of Modern Physics, Chinese Academy of Sciences, 730000 Lanzhou, China}

\affiliation{\it {}$^{14}$
 Instituto de Estructura de la Materia -- CSIC, 28006 Madrid, Spain}

\affiliation{\it {}$^{15}$
 Lanzhou University, 730000 Lanzhou, China}

\affiliation{\it {}$^{16}$
 IGFAE, Universidade de Santiago de Compostela, 15782 Santiago de Compostela, Spain}

\affiliation{\it {}$^{17}$
 Jo\v{z}ef Stefan Institute, 1000 Ljubljana, Slovenia}

\affiliation{\it {}$^{18}$
 Faculty of Mathematics and Physics, University of Ljubljana, 1000 Ljubljana, Slovenia}

\affiliation{\it {}$^{19}$
 Institute of Science Tokyo, Meguro, 152--8550 Tokyo, Japan}

\affiliation{\it {}$^{20}$
 Institut f\"{u}r Kernphysik, Forschungszentrum J\"{u}lich, 52425 J\"{u}lich, Germany}

\affiliation{\it {}$^{21}$
 Institut f\"{u}r Kernphysik, Technische Universit\"{a}t Darmstadt, 64289
  Darmstadt, Germany}
  
\affiliation{\it {}$^{22}$
 Jagiellonian University, 30--348 Krak\'{o}w, Poland}

\affiliation{\it {}$^{23}$
INFN, Laboratori Nazionali di Frascati, 00044 Frascati, Roma, Italy}

\affiliation{\it {}$^{24}$
 Institute for Medical Physics and Radiation Protection, TH Mittelhessen University of Applied Sciences, 35390 Gie\ss en, Germany}

\affiliation{\it {}$^{25}$
 National Superconducting Cyclotron Laboratory, Michigan State University, MI 48824 East Lansing, USA}

\affiliation{\it {}$^{26}$
 CITENI, Campus Industrial de Ferrol, Universidade da Coru\~na, 15403 Ferrol, Spain}

\affiliation{\it {}$^{27}$
 Variable Energy Cyclotron Centre, 1/AF--Bidhan Nagar, 700064 Kolkata, India}

\affiliation{\it {}$^{28}$
 Uppsala University, 75220 Uppsala, Sweden}

\affiliation{\it {}$^{29}$
 Helmholtz Forschungsakademie Hessen f\"{u}r FAIR (HFHF), GSI Helmholtzzentrum
  f\"{u}r Schwerionenforschung, Campus Gie{\ss}en, 35392 Gie{\ss}en, Germany}

\affiliation{\it {}$^{30}$
 High Energy Accelerator Research Organization (KEK), Tsukuba, 305--0801 Ibaraki, Japan}

\affiliation{\it {}$^{31}$
 Department of Physics, Saitama University, 338--8570 Saitama, Japan}

\affiliation{\it {}$^{32}$
 Peking University, 100871 Beijing, China}

\collaboration{$\eta$-PRiME/WASA-FRS/Super-FRS Experiment Collaboration}

\begin{abstract}
  The missing mass of the  $\C{12}(p,d)$ reaction has been measured near the $\eta'$-meson emission threshold
  in coincidence with a high-momentum proton to selectively collect $\eta'$-$\C{11}$ mesic nucleus
  formation events at GSI, Germany. 
  A 2.5 GeV proton excites a carbon nucleus
  to form an $\eta'$-mesic nucleus emitting a deuteron forward
  with an energy of $\sim 1.6$ GeV. The deuteron is momentum-analyzed by the Fragment Separator 
  used as a high-resolution spectrometer to
  deduce the excitation energy of the residual system. 
  The large-acceptance detector WASA surrounding the target
  identifies high-momentum protons emitted in the decay of the $\eta'$-mesic nucleus.
 The measured semi-exclusive spectrum exhibits structures below the threshold 
 though the statistical significance is limited. 
 The spectrum is fitted by theoretically calculated spectra varying 
 optical-potential parameters
 of the $\eta'$-nucleus interaction.
  The analysis results indicate $\eta'$-mesic nuclei formation for
the real potential depth of 
$\sim -61$ MeV
with a local statistical significance of $3.5 \sigma$
  and, taking into account the look-elsewhere effect, 
 a global significance of $2.1 \sigma$.   
\end{abstract}
\maketitle

Nine members compose the lowest-mass multiplet of pseudo-scalar mesons, namely $\pi^\pm, \pi^0$, $K^\pm,
K^0, \overline{K^0}, \eta$ and $\eta'$. Their masses are widely distributed from 135 to 958 MeV/$c^2$
reflecting the broken symmetry of the underlying vacuum~\cite{Nagahiro:2013fk,Bass2020}.
    The nonet includes massless Nambu-Goldstone bosons~\cite{PhysRev.117.648} generated in the dynamical breakdown of
    the flavor SU(3) chiral symmetry of the vacuum in the low-energy region of the quantum chromodynamics (QCD), and
    the nine members are expected to be mass-degenerate if the chiral symmetry were to hold.
    Investigations of the mass distribution in a chiral-symmetry-restored environment
    provide information on the mass generation mechanisms
    and the non-trivial structure of the vacuum in the evolution of the universe.
    We hereby focus on $\eta'$, which has an exceptionally large mass of 958 MeV/$c^2$
    in the nonet. The origin of the large
    mass is yet to be explained and has been known as the $\eta$ problem since
    1970's~\cite{PhysRevD.11.3583}.
    According to modern theories, the large mass is attributed to coupling of
    the chiral condensate $\qq$ and the axial U(1) quantum anomaly of
    QCD~\cite{Witten:1979fk} in relation to 
    the anomalous gluon dynamics~\cite{Bass2020}.
    Due to partial restoration of the chiral symmetry,
    an order parameter, the vacuum expectation value of the chiral condensate $|\qq|$,
    and consequently also the $\eta'$ mass are expected to be lowered in the nucleus~\cite{Nishi/2023_NP19}.

    Since $|\qq|$ is reduced by $40\pm 3\%$ at the nuclear saturation density~\cite{Nishi/2023_NP19},
    a large $\eta'$ mass reduction $|\Delta m|$ $(\Delta m<0)$ is expected. The mass reduction
    infers the existence of a scalar attractive potential of the same magnitude.
    A complex 
    potential  $U(r) = (V_0 + i W_0)\rho(r)/\rho(0)$ is assumed using 
    two real parameters, $V_0 =\Delta m$ and $W_0$, 
    where $\rho(r)$ denotes the nuclear density~\cite{Nagahiro:2013fk}.
    Many theories predict 
    relatively different values of the potential parameters while a few experiments set constraints. The
    Nambu--Jona-Lasinio model predicts a strongly attractive potential of 
    $V_0=-150$ MeV~\cite{PhysRevD.71.116002,PhysRevC.74.045203}. The linear sigma model~\cite{PhysRevC.88.064906} and
    the quark meson coupling (QMC) model~\cite{BASS2006368, Cobos_Tsushima_PRC2024}
    predict shallower potentials in a range of [$-80$, $-37$]~MeV.
   Recent photoproduction experiments yield $V_0= -[39 \pm 7 ({\it stat}) \pm 15({\it syst})]$ MeV by measuring excitation functions and momentum distributions~\cite{Nanova2013417,PhysRevC.94.025205,Nanova18}, 
   and $W_0= -[13 \pm 3 ({\it stat}) \pm 3 ({\it syst})]$ MeV
   by determining transparency ratios~\cite{NANOVA2012600,Friedrich/Friedrich2016aa}. 

    Measurements of quantum bound states of an $\eta'$ in a nucleus
    provide accurate and robust information on the $\eta'$ properties in nuclear matter.
    Large $|V_0|$ and small $|W_0|$ may accommodate discrete quantum levels of bound $\eta'$
    in the nucleus, $\eta'$-mesic nuclei~\cite{PhysRevC.85.032201}.
    There have been experiments to search for the bound states using an
    inclusive $\C{12}(p,d)$ reaction~\cite{PTP.128.601,PhysRevLett.117.202501,PhysRevC.97.015202}
    and a semi-exclusive $\C{12}(\gamma,p)$ reaction with $\eta p$ tagging~\cite{PhysRevLett.124.202501,PhysRevLett.126.019201}.
    However, these states have not been discovered so far. Their observation would be a manifestation of a meson-nucleus system solely bound by the strong interaction.
    With potential parameters $|W_0| \ll |V_0|$, 
    $\eta'$ becomes a promising candidate for the experimental search of meson-nucleus bound states.
    
    In the preceding experiment to search for the $\eta'$-mesic nuclei, the $\C{12}(p,d)$ reaction was employed, where
    the incident proton excites the carbon nucleus near the $\eta'$ emission threshold. 
    The measured inclusive missing-mass spectra of the $\C{12}(p,d)$ reaction 
    present a structureless continuum mainly due to the quasi-free meson production ``background" processes
    in the $(p,d)$ reaction~\cite{PhysRevLett.117.202501,PhysRevC.97.015202}. 
    The theoretically calculated signal cross sections~\cite{Nagahiro:2013fk} are smaller than the measured 
    inclusive cross section of 4.9--5.7 $\mu$b/(sr$\cdot$MeV) by 2--3 
    orders of magnitude.
    Although one has to allow for systematic uncertainties
    originating in the elementary $\eta'$ production cross section of the $n(p,d)\eta'$ reaction~\cite{Grishina_PLB2000},
    due to the high statistics, constraints are set on potential parameters as in Fig.~12 of Ref.~\cite{PhysRevC.97.015202}.

    In the present experiment, events of the $\eta'$-mesic nuclei formation are tagged
    in the missing-mass spectroscopy of the $\C{12}(p,d)$ reaction via their decays
    to reduce the background. 
    Since the $\eta'$-mesic nuclei are produced with smaller momentum in the laboratory frame, the decay particles
    are emitted almost isotropically while most of the background processes emit particles forward.
    Thus, the signal-to-background ratio (S/B) is improved by tagging decay particles in the backward region.
    Possible decay channels are predicted theoretically, namely,
    1) $\etap N \rightarrow M N$,
    2) $\etap NN \rightarrow NN$, where $N$ and $M$ denote a nucleon and a meson, respectively ~\cite{Nagahiro2012PLB}.
    The $M N$ decay channels may be $\eta N$, $\pi N$, $K\Lambda$ and $K\Sigma$.
    The branching ratio Br$_{M\! N}$:Br$_{N\! N}$ is derived to be
    $\sim 1:1$ for the case of $|a_{\eta'N}| = 0.5$ fm~\cite{Nagahiro2012PLB}.
    Among them, the channel 2 ``two nucleon absorption'' is promising, because background
    processes scarcely generate such an energetic nucleon of $\sim 1$ GeV/$c$ in the backward region
    as in the signal process~\cite{Ikeno_arxiv_2024}.

    Simulated semi-exclusive
    spectra expected for 3 days of data taking present near-threshold structures 
    for larger $|V_0|$ and smaller $|W_0|$.
The theoretical formation spectra with selecting the two-nucleon absorption channel 
    were calculated for differently assumed optical potential 
         $U(r) = (V_0 + i W_1)\rho(r)/\rho(0) + i W_2 (\rho(r)/\rho(0))^2$
     by using the Green's function method \cite{Nagahiro:2013fk, Ikeno_arxiv_2024}.
   The absorption term of the potential is decomposed into $W_1$ and $W_2$ representing single- and two-nucleon absorption, respectively, and a case of $W_0/2 \simeq W_1 = W_2$ is assumed for simplicity. The expected spectral shapes barely change for different assumptions of $W_1/W_2$ ratios.
 A nuclear transport simulation JAM~\cite{Nara_PRC2000_JAM}
    has deduced a factor of
    $\sim 100$ S/B improvement by tagging the protons within the momentum range $[0.7,1.2]$~GeV/$c$
    in the backward region $\theta_p > 90^{\circ}$~\cite{Ikeno_arxiv_2024}.
    Note that the simulation does not include
    effects of the short-range correlated pairs of nucleons in the carbon nuclei so that 
    the S/B in reality may be slightly smaller.
The results of the simulated semi-exclusive spectra demonstrate 
the feasibility of observing a near-threshold peak structure when $|V_0| \gg |W_0|$. 
    
\begin{figure}[hbtp]
  \centering{
    \vskip0cm
    \hskip0cm
  \includegraphics[width=8.0cm]{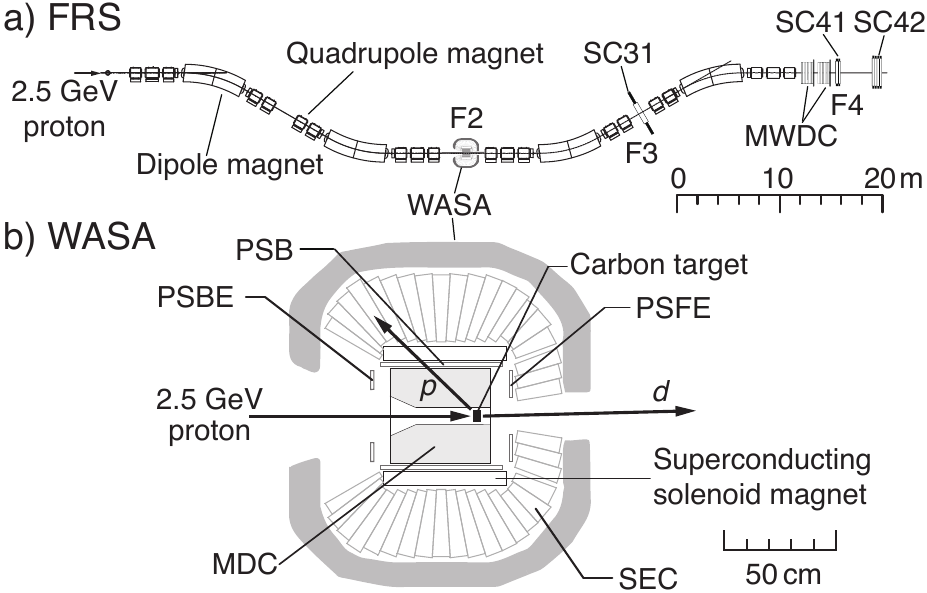}
    \vskip0cm
    \caption{Schematic setup of a) FRS and b) WASA.
      WASA was installed at F2, 
      where a 2.5 GeV proton beam impinged on
      a $\C{12}$ target for the $(p,d)$ reaction. 
      The emitted $d$ is momentum-analyzed
      in the F2-F4 section of the FRS. 
    Protons of $\sim 1$ GeV/$c$ are tagged by WASA. 
  }
\vskip-3mm
  \label{Fig:setup}
}
\end{figure}
High-resolution missing-mass spectroscopy 
of the $\C{12}(p,d)$ reaction was performed at the Fragment Separator (FRS)~\cite{FRS_Geissel_92} in GSI, Darmstadt,
in combination with tagging $\sim 1$ GeV/$c$ protons.
Figure~\ref{Fig:setup} schematically depicts the experimental setup.
A $2497.4 \pm 4.9$ MeV proton beam 
accelerated by the SIS-18 synchrotron 
impinged on a $4.01 \pm 0.02$ g/cm$^2$ C(nat) target 
at F2.
The beam intensity was $\sim 5 \times 10^8$/s, and the spill length and cycle were
10 s and 11 s, respectively. The intensity was continuously monitored 
with an accuracy of 
$\sim 5\%$. 
The beam spot size was about 1~mm ($\sigma$) in both horizontal and vertical directions and was
stable within $\pm 0.5$ mm throughout the experiment. 
$1.1 \times 10^{7}$ events of the $\C{12}(p,d)$ reaction
were accumulated in a data-taking period of 3 days.
The 
WASA central detector~\cite{WASA1_Bargholtz_NIMA2008} was installed
at F2 to tag the $\sim$ 1 GeV/$c$ protons.

The forward emitted deuteron of $\sim2.8$~GeV/$c$
was momentum-analyzed in the F2-F4 section of the FRS 
to deduce the excitation energy ($E_\mathrm{ex}$) of the $\C{11}$ nucleus
relative to the $\etap$ emission threshold ($E_0$) in the range of $[-70,50]$~MeV. 
Multi-wire drift chambers (MWDCs) were installed at F4 to measure the deuteron tracks.
The momentum dispersion and magnification at F4 were 42~mm/\% and 1.3, respectively. 
The resultant spectral resolution is $\sigma_{\rm ex}=1.6\pm0.1$~MeV,
including contributions from energy straggling in the target and momentum spread of the primary proton beam. 
The optical mode had a momentum acceptance of $\sim \pm 2.5\%$ and angular
acceptance of 
$2$~msr at the central momentum. 

The $E_{\rm ex}$ scale  was calibrated 
with an accuracy of 2.5~MeV 
by measuring
D$(p,d)p$ elastic scattering 
on a $4.03 \pm 0.02$~g/cm$^2$ deuterated polyethylene 
target. 
The FRS setting was kept identical to the production runs, and
a 1622.3$\pm$3.0~MeV proton beam 
was employed to produce
nearly mono-energetic deuterons with a momentum of 2820.4$\pm$3.5 MeV/$c$
after energy loss in the materials.

For the acceptance corrections of the spectrometer, 
the inclusive $\C{12}(p,d)$ spectrum has been used, which has no distinct structures
in the measured range $E_{\rm ex}-E_0 \sim [-70,50]$~MeV being consistent with the preceding experiment~\cite{PhysRevLett.117.202501,PhysRevC.97.015202}.
The acceptance corrections are made so as to reproduce the preceding results
of the inclusive measurement expressed in the
third-order polynomial 
of Eqn.~(8) in Ref.~\cite{PhysRevC.97.015202}.  
The above derived acceptance inherently contains bin-by-bin
efficiency correction of the $E_{\rm ex}$ spectrum.  

Three sets of plastic scintillation counters SC31 at F3 and SC41 and SC42 at F4
identified deuterons by measuring the time of flight (TOF). 
A typical counting rate at SC41 
was 30 kHz,
where 99.5\% 
was due to background protons.
The entire data acquisition system was triggered 
by a SC31-SC41 TOF-based coincidence.
The TOF trigger suppressed 99.3\% of the background protons, while $\sim$100\% deuterons were selected. 
Background-free identification of deuterons was achieved 
by further analysis of the SC41-SC42 TOF.

WASA consists of a 1 T superconducting solenoid magnet 
\cite{Ruber_2003},
a cylindrical tracking detector (MDC),
a plastic scintillator barrel (PSB) and arrays (PSFE, PSBE),
and a scintillator electromagnetic calorimeter (SEC).
The carbon target was installed 
at a position 15 cm downstream of the center of WASA
so that the MDC covers the polar angle $\theta_p$ of
$[73,151]$ degrees. Details of these detectors are described elsewhere~\cite{WASA1_Bargholtz_NIMA2008, SEKIYA2022166745, Acta2023}.

The WASA analysis includes a charge over momentum ($q/P$) measurement with 
a trajectory reconstruction by the MDC and 
hit timing ($t_{\rm PSB}$) and 
energy loss ($\Delta E$) measurements 
by the PSB.
An elastic-arm approach~\cite{Ohlsson_CompPhys_1992_EAA} was applied  
to find track candidates out of the possible hit wires in the MDC.
The hit candidates were fitted with a Kalman-filter algorithm 
to reconstruct the trajectories and $q/P$ using the GenFit package~\cite{Genfit_NIMA}.
Waveforms of the PSB signals 
were analyzed 
to deduce $t_{\rm PSB}$ and $\Delta E$ \cite{PSB_arxiv_2025}. 
The charged-particle velocity $\beta$
was evaluated in two different ways, namely $\beta_\mathrm{TOF}$ and $\beta_{\Delta E}$.
The former was evaluated 
by the time-of-flight from the target to the PSB,
where the time at the target was estimated using the SC41 hit timing 
with ion-optical corrections.
The latter was calibrated by the correlation between $\Delta E$ and $\beta_\mathrm{TOF}$.

\begin{figure}[hbtp]
   \includegraphics[width=8.5cm]{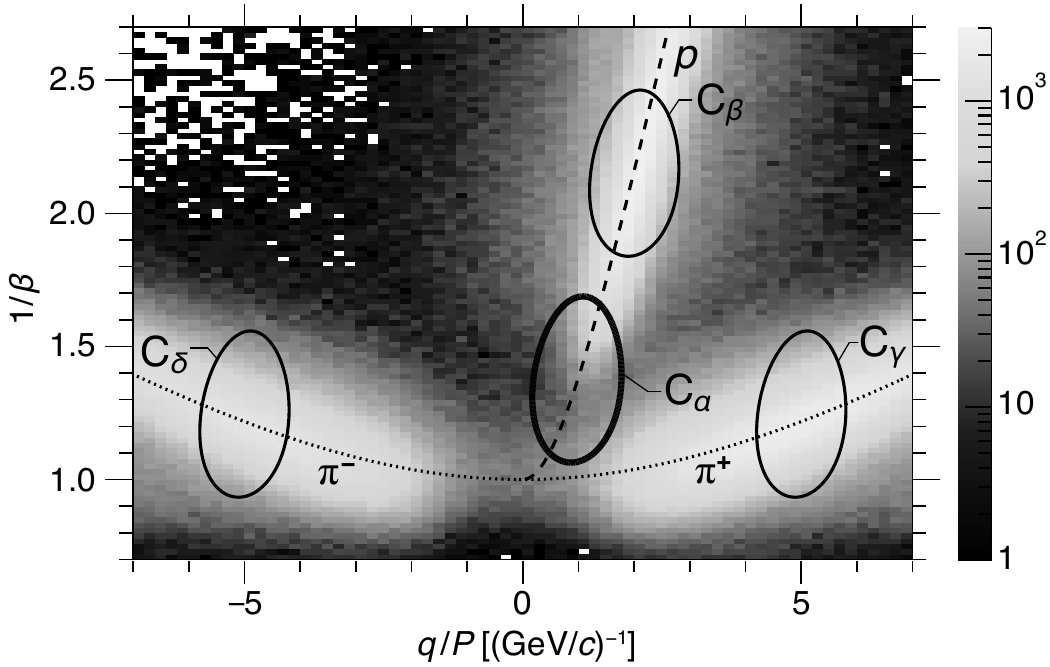}
  \caption{The particle identification 
  in WASA. The abscissa is $q/P$ measured by the MDC and the ordinate is $1/\beta$ obtained by the weighted average of $1/\beta_\mathrm{TOF}$ and $1/\beta_{\Delta E}$. The dotted line indicates the theoretical line for $\pi^{\pm}$ and the dashed line for proton.  
  C$_{\alpha}$ 
  marks the signal regions for energetic protons emitted in the decay of the $\eta'$-mesic nucleus. 
Background dominant regions gating on low-energy protons, $\pi^+$, and $\pi^-$ are marked by C$_{\beta}$, C$_{\gamma}$, and C$_{\delta}$, respectively.
}
\vskip-5mm
  \label{Fig:wasa_analysis}
\end{figure}

Figure \ref{Fig:wasa_analysis} shows the particle identification in WASA by using the MDC and PSB information.
The abscissa is the deduced $q/P$ and the ordinate is $1/\beta$ obtained by a weighted average of $1/\beta_\mathrm{TOF}$ and $1/\beta_{\Delta E}$ taking 
into account the resolutions.
The charged particles, 
$\pi^{\pm}$ and $p$, are well identified.
A Monte Carlo simulation was performed using the Geant4 package~\cite{Agostinelli_NIMA_2003_Geant4},
to evaluate various efficiencies and resolutions.
The detection and analysis efficiency of WASA is evaluated to be 72\% and the resolutions of $q/P$, $1/\beta_\mathrm{TOF}$, 
and $1/\beta_{\Delta E}$  to be 27\%, 11\%, and 8.2\%, respectively, for 1 GeV/$c$ protons emitted at $\theta_p=90^{\circ}$.

Efficient 
selections of the signal contributions have been achieved 
by choosing energetic protons in the C$_\alpha$ region presented in Fig.~\ref{Fig:wasa_analysis}, which
reflect the expected signal proton energy distribution~\cite{Ikeno_arxiv_2024}. Intra-nuclear transport simulations have been
performed by using the JAM~\cite{Nara_PRC2000_JAM} and GiBUU~\cite{GiBUU_review} 
packages to
obtain 
realistic signal proton distributions in the two-nucleon absorption decay channel. 
Both simulations have consistently yielded a longer tail towards the lower energy region due to the final-state interactions,
which slightly lower the efficiency of the C$_\alpha$ cut. 
Applying the C$_{\alpha}$ cut and $\theta_p > 90^{\circ}$, 
 the number of $(p,d)$ reaction events
is reduced by a factor of 446, which agrees with the expected
reduction factor of $\sim 100$. 
For better comparison, three ``sideband" cut conditions in the background-dominant regions
are prepared, namely, 
C$_\beta$ for lower momentum protons, C$_\gamma$ for $\pi^+$, and
C$_\delta$ for $\pi^-$.

\begin{figure}[hbtp]
\centering{
    \includegraphics[width=8.5cm]{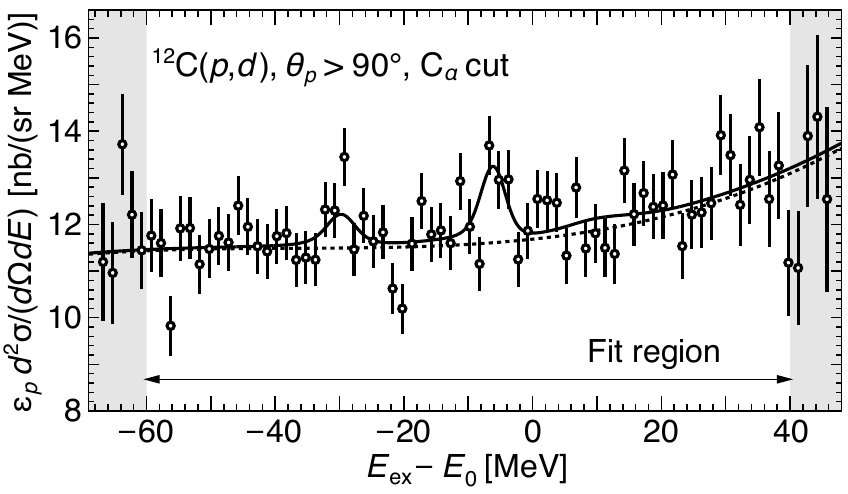} 
  \caption{Measured excitation spectrum of $\C{12}(p,d)$ reaction near the $\eta'$ emission threshold $E_0$
  with high-momentum backward proton selection by the C$_\alpha$ cut and $\theta_p>90^\circ$.  
  A fit has been performed 
  using the sum of $f_{\rm sig}$ and $f_{\rm ped}$, as shown by 
  the solid curve for $(V_0,W_0) = (-62,-2)$~MeV. The $f_{\rm ped}$ is shown by the dotted curve. 
      }
  \label{Fig:spectrum}
\vskip -3mm
}
\end{figure}

Figure~\ref{Fig:spectrum} shows the excitation spectrum 
with the C$_{\alpha}$ cut in Fig.~\ref{Fig:wasa_analysis} and $\theta_p>90^\circ$.
The abscissa is $E_{\rm ex}$ relative to the $\eta'$ emission threshold $E_0 = 957.78$ MeV.
The ordinate is the double differential cross section $(d^2\sigma/d\Omega dE)$ of the $\C{12}(p,d)$ reaction
multiplied by the tagging probability $\varepsilon_p$ of the high-momentum backward protons in the semi-exclusive measurement
per inclusive $(p,d)$ reaction.
Two structures are observed near the $\eta'$
emission threshold at 
$E_{\rm ex}-E_0 = -30$ and $-6$ MeV
standing on a smooth pedestal. 
This 
difference of $\sim24$~MeV is consistent with expected level spacings of the $\eta'$-$^{11}$C 
bound states~\cite{Nagahiro:2013fk, Ikeno_arxiv_2024}.
The same analysis with sideband cut conditions ($C_\beta$, $C_\gamma$ and $C_\delta$) and $\theta_p > 90^\circ$
generates spectra consistent with a smooth pedestal (see Supplemental Material).

The possible formation of $\eta'$-mesic nuclei is investigated
by fitting the spectrum with a sum of the signal contribution $f_{\rm sig}$ and a smooth pedestal component $f_{\rm ped}$. 
The $f_{\rm sig}$ pattern is given by the resolution $\sigma_{\rm ex}$-folded theoretical cross section for a set of $(V_0,W_0)$
branching to the two-nucleon absorption channel $f^\sigma_{(V_0,W_0)}$, 
multiplied by three factors, namely, 1) the expected number of protons $\bar{N_p}=0.342$ emitted 
in $\theta_p > 90^{\circ}$ 
from a two-nucleon absorption decay, 
2) the detection and reconstruction efficiency of WASA for these protons $\omega_\mathrm{WASA}=0.561$,
including the selection with the $C_\alpha$ cut,
and 3)
a scale parameter $\mu$ so that $f_{\rm sig} = \mu\,\bar{N_p}\,\omega_\mathrm{WASA}\,f^\sigma_{(V_0,W_0)}$.
Note that the branching ratio $\mathrm{Br}_{N\!N}$ is reflected in the theoretical spectrum $f^\sigma_{(V_0,W_0)}$.
$\bar{N}_p$ takes into account all possible combinations of protons and neutrons in a $^{11}$C nucleus
as well as the survival probability due to the final-state interaction
evaluated by JAM.
The theoretical spectra $f^\sigma_{(V_0,W_0)}$ are calculated on a $(V_0, W_0)$ grid with a step of 2 MeV 
in the ranges of $V_0$ in $[-150, 0]$ MeV and $W_0$ in $[-30, -2]$ MeV.
Here, a third-order polynomial of $E_{\rm ex}$ is assumed for the $f_{\rm ped}$ providing sufficient degrees of freedom,
which fitted the inclusive spectrum well in the preceding experiment even with much higher statistics and over a 
wider $E_{\rm ex}$ range~\cite{PhysRevC.97.015202}.
The spectrum is fitted 
in the range of $[-60,40]$ MeV
treating $\mu$ and the polynomial coefficients as free parameters
with the constraint $0 \leq \mu$ 
for each $(V_0,W_0)$ on the grid.
The best fit result is obtained for
$(V_0,W_0) = (-62, -2)$ MeV, shown 
by the solid
curve and its pedestal component as the dotted curve.
The fit yields 
$\chi^2/\mathrm{(n.d.f.)} =  65.76 / 62 $ and $\mu = 1.43 \pm 0.42$.
In contrast, the pedestal-only fit yields 
$\chi^2/{\rm (n.d.f.)} = 77.36 / 63$, resulting in a $\chi^2$ difference of 
$\Delta \chi^2 = 11.60$.
The rise of $f_{\rm ped}$ above threshold is probably due to the
onset of quasi-free $\eta'$ production,
which requires further experimental and theoretical investigations~\cite{Bruns19,Sakai23}.

\begin{figure}[hbtp]
\centering{
   \includegraphics[width=8.6cm]{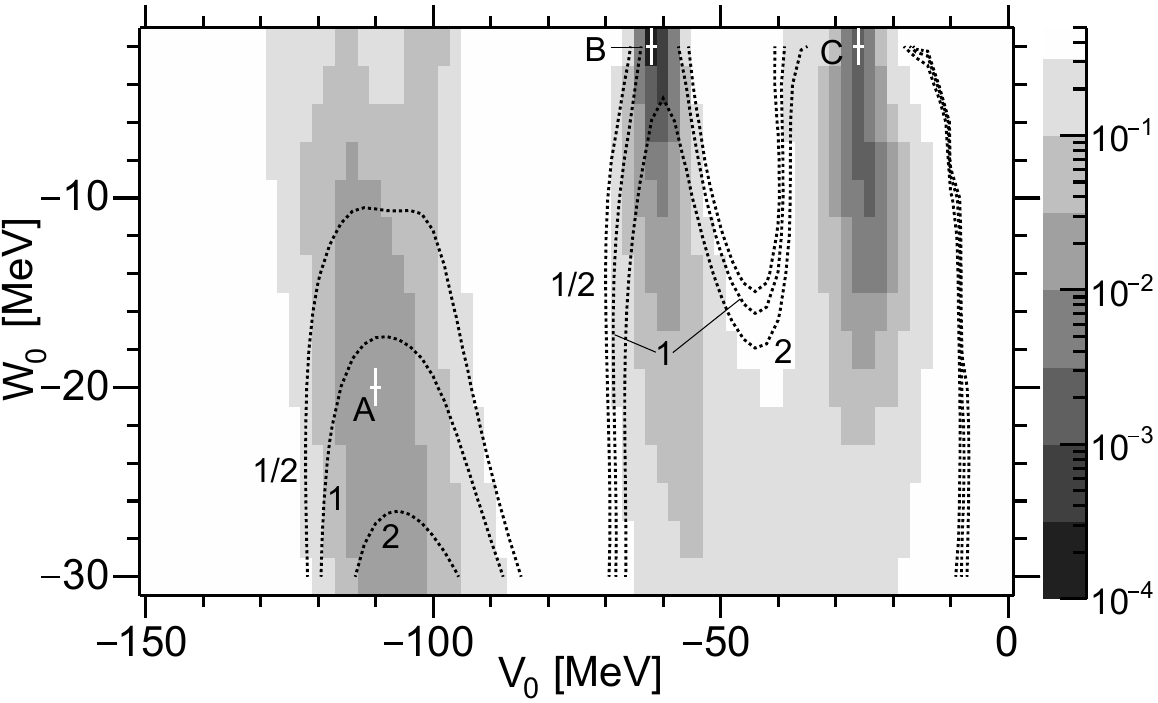}
  \caption{Local $p$-values presented by the gray scale on the $V_0$-$W_0$ plane calculated for each 2 MeV bin.
Local minima are found for three points  
A:$(V_0,W_0)=(-110,-20)$, B:$(-62,-2)$ and C:$(-26,-2)$ MeV.
The fitted $\mu$ is overlaid as the dotted contours. 
    }
\vskip-3mm
  \label{Fig:vw}
}
\end{figure}

In order to evaluate the $p$-value and statistical significance of the signal with respect to the pedestal-only hypothesis,
$\Delta \chi^2$ 
is analyzed for each $(V_0, W_0)$ using Monte-Carlo simulations.
The obtained local $p$-values on the 
($V_0, W_0$) grid are presented in Fig.~\ref{Fig:vw}.
Local minima are found with $p$-values of $1.3\times 10^{-2}$, $2.6\times 10^{-4}$, and $2.1 \times 10^{-3}$
at (A) $(V_0,W_0) = (-110,-20)$, (B) $(-62,-2)$, and (C) $(-26,-2)$ MeV, 
respectively.
The local $p$-value for the case A is by an order of magnitude 
larger than the others 
making it less likely.
Figure~\ref{Fig:vw} also presents the fitted $\mu$ in the dotted contours.
The fitted $\mu$ must be consistent with unity within the uncertainties of the expected signal cross sections,
making the case C unlikely due to its too large value 
of $\mu \sim 13.8 \pm 4.9$.

The case B exhibits the smallest local $p$-value of $2.6\times 10^{-4}$
corresponding to the statistical significance of 3.5$\sigma$
and a consistent value of $\mu = 1.43 \pm 0.42$ as obtained in the fit of Fig.~\ref{Fig:spectrum}.
This local significance
leads to a global $p$-value of $1.7 \times 10^{-2}$ and significance of 2.1$\sigma$
when the look-elsewhere effect~\cite{Lyons_look_elsewhere} is taken into account.
In this case, structures observed at $E_{\rm ex}-E_0 = -30$ and $-6$ MeV are assigned to excitation of the
$\eta'$-mesic nuclei in the $1s$ and $2p$ orbitals, respectively \cite{Nagahiro:2013fk,Ikeno_arxiv_2024}.
An analysis employing an unbinned maximum likelihood method has been also performed, yielding nearly identical results.

Finally, the $\chi^2$ evaluation around point B indicates
$V_0= -61 \pm 1 (stat.) \pm 5 (syst.)$ MeV after interpolation between the grids.
An upper limit of $|W_0| \lesssim 10$~MeV (68\% C.L.) has been deduced.
Systematic uncertainties have been evaluated by repeating the analysis 
by varying the absolute calibration of $E_{\mathrm{ex}}$ and the resolution $\sigma_{\rm ex}$ within the errors
and by trying alternative choices of the pedestal functions. 
The above results agree with the photoproduction experiment within the uncertainties~\cite{NANOVA2012600,Friedrich/Friedrich2016aa}.
It is in line with QMC model calculations~\cite{Cobos_Tsushima_PRC2024} 
and with the chiral unitary approach~\cite{Nagahiro2012PLB}
but in conflict with recent calculations~\cite{Kumar2025_CPC,Gal2025}.

In conclusion, spectroscopy of 
$\C{12}(p,d)$ reaction has been performed near the $\eta'$ emission threshold,
measuring the emitted particles around the target. The observed semi-exclusive spectrum
measured in coincidence with a high-momentum proton displays structures 
near $E_{\rm ex}-E_0 = -30$ and $-6$ MeV.
The results indicate the first direct detection of $\eta'$-mesic nuclei, which provide
information on the meson properties in a high-density nuclear medium.
The statistical analysis yields an $\eta'$ mass modification of 
$\Delta m = -61 \pm 1 (stat.) \pm 5 (syst.)$ MeV$/c^2$
at the nuclear saturation density.
In the future, the significance will be enhanced for a wider $E_{\rm ex}$ range enabling robust conclusions and 
unambiguous determination of the quantum levels. Increasing the acceptance of the tagging detector,
observation of two back-to-back two nucleons will become possible.
In addition, different decay channels will be measured such as $K\Lambda$. 
At J-PARC or FAIR, pion beams can be used to exploit larger
formation cross sections. We also aim at measurement of the elementary $\eta'$ production cross section.

The experimental results reported here in the context of FAIR Phase-0
were obtained in the experiment S490, which was performed at the FRS 
at the GSI Helmholtzzentrum f\"ur Schwerionenforschung, Darmstadt (Germany) 
in the context of FAIR Phase-0.
The authors thank the staffs of FZJ and GSI for transport,
installation, and operation of WASA and FRS.
This work is partly supported by
JSPS Grants-in-Aid for Scientific Research (A) (No.~JP24H00238), (B) (No.~JP18H01242)
and for Early-Career Scientists (No.~JP20K14499), 
JSPS Fund for the Promotion of Joint International Research
(Fostering Joint International Research (B)) (No.~JP20KK0070).
The authors would like to acknowledge supports
from the SciMat and qLife Priority Research Areas budget under the program Excellence Initiative-Research University at the Jagiellonian University, 
from Proyectos I+D+i 2020 (ref: PID2020-118009GA-I00), 
from the program ‘Atracci\'{o}n de Talento Investigador’ of the Community of Madrid (Grant 2019-T1/TIC-131), 
from the Regional Government of Galicia under the Postdoctoral Fellowship Grant No.~ED481D-2021-018, 
from the MCIN under Grant No.~RYC2021-031989-I, 
from the ExtreMe Matter Institute EMMI at the GSI Helmholtzzentrum f\"{u}r Schwerionenforschung, Darmstadt, Germany,
and from the European Union’s Horizon 2020 research and innovation programme (Grant No.~824093).

\onecolumngrid
\clearpage
\begin{center}
\boldmath{ \large \bf
 Supplemental Material: Excitation Spectra of the $\C{12}(p,d)$ Reaction near the $\eta'$-Meson Emission Threshold Measured in Coincidence with High-Momentum Protons}
\end{center}
\vspace{2mm}
\twocolumngrid
\renewcommand{\thefigure}{S\arabic{figure}}
\setcounter{figure}{0} 

{\noindent \bf{{\boldmath{Excitation-energy spectra with sideband cuts}}}}  \vskip2mm
   
\begin{figure}[t]
  \centering{
    \vskip0cm
    \hskip0cm
  \includegraphics[width=7.5cm]{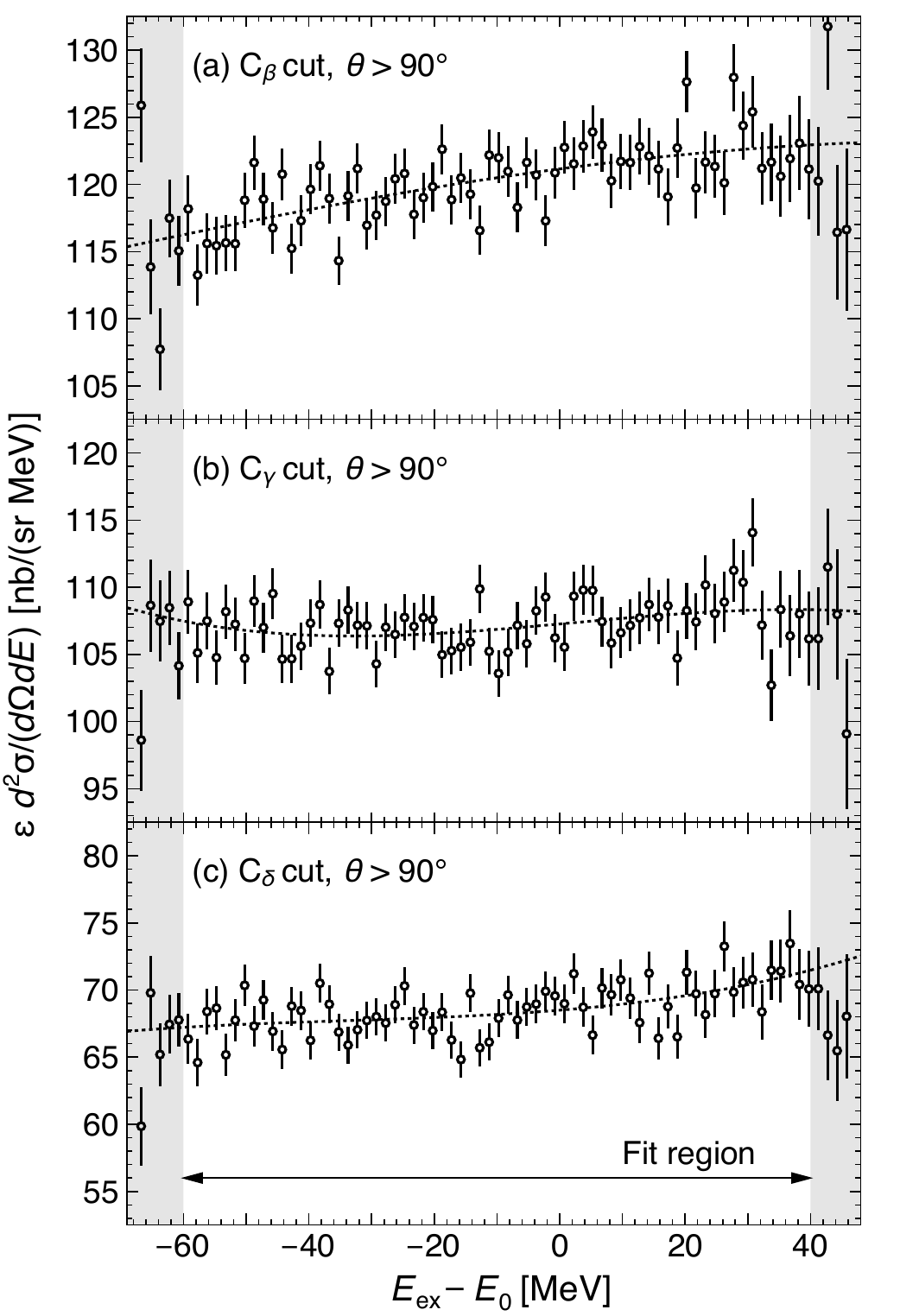}
    \vskip0cm
    \caption{Measured excitation spectra of 
    $^{12}$C($p$,$d$) reaction 
    with selections of backward-emitted ($\theta > 90^{\circ}$) particles in WASA
    by the sideband cuts,
    (a) $\mathrm{C}_\beta$ cut on lower momentum protons,
    (b) $\mathrm{C}_\gamma$ cut on $\pi^{+}$, 
    and (c) $\mathrm{C}_\delta$ cut on $\pi^{-}$.
    Fits have been performed in the region indicated by the arrows 
    using pedestal functions given by a third-order polynomial, as shown by the dotted curves.
  }
\vskip-3mm
  \label{Fig:suppl_fig1}
\vskip-10mm
}
\end{figure}

Here, we provide spectra of $^{12}$C($p$,$d$) reaction
in coincidence with backward-emitted ($\theta>90^{\circ}$) particles in WASA 
in the background-dominated sideband regions, defined by the cuts in Fig.~2 of the main article.
Figure~\ref{Fig:suppl_fig1} shows the measured spectra of the excitation energy ($E_\mathrm{ex})$ 
near the $\eta^\prime$ emission threshold $E_0 = 957.78$~MeV
in the $^{12}$C($p$,$d$) reaction 
with sideband selections of (a) low-momentum protons by the $\mathrm{C}_\beta$ cut,
(b) $\pi^{+}$ by the $\mathrm{C}_\gamma$ cut, and (c) $\pi^{-}$ by the $\mathrm{C}_\delta$ cut. 
The ordinate represents the double differential cross section  ($d^2\sigma/d\Omega dE$) 
of the $^{12}$C($p$,$d$) reaction multiplied by the tagging probability $\varepsilon$ of 
each corresponding particle in the semi-exclusive measurement per inclusive ($p$,$d$) reaction.  
The obtained spectra exhibit a smooth behavior with no significant peak structures.
The spectra have been fitted in the range of [$-60$,$40$]~MeV
using pedestal functions expressed by a third-order polynomial, as indicated by the dotted curves,
yielding $\chi^2/\mathrm{(n.d.f.)} = 64.87/63$, $56.41/63$, and $67.62/63$ for the $\mathrm{C}_\beta$, $\mathrm{C}_\gamma$, and $\mathrm{C}_\delta$ cuts, respectively.

In order to investigate possible signal contributions, 
the same analysis procedure as described in the main text 
has been applied to the spectra with the sideband selections. 
The $\Delta \chi^2$ values, the difference between the $\chi^2$ of the pedestal-only fit and
the minimum $\chi^2$ of the signal-and-pedestal fit obtained from the scan over the ($V_0$, $W_0$) grid,
are $2.03$, $2.07$, and $5.40$ for the $\mathrm{C}_\beta$, $\mathrm{C}_\gamma$, and $\mathrm{C}_\delta$ cuts, respectively. 
The corresponding global $p$-values under the pedestal-only hypothesis amount to about 0.9, 0.9, and 0.3, respectively, 
showing no indication of signals associated with the formation of $\eta^\prime$-mesic nuclei in the sideband spectra.

\newpage

\end{document}